\documentclass[runningheads]{llncs}

\usepackage[T1]{fontenc}

\usepackage{graphicx}

\usepackage[
	colorlinks,
	allcolors=blue
]{hyperref}
\usepackage{color}

\usepackage[backend=biber,
	style=lncs
]{biblatex}
\begin{document}

\title{Appending Data to Blockchain is not Sufficient for Non-repudiation of Receipt}

\titlerunning{Appending to Blockchain not Sufficient for Non-repudiation of Receipt}

\author{Valentin Zieglmeier\orcidID{0000-0002-3770-0321}}

\authorrunning{V. Zieglmeier}

\institute{%
	Technical University of Munich, Munich, Germany\\
	\email{valentin.zieglmeier@tum.de}%
}

\maketitle

\begin{abstract}
Exchanging data while ensuring non-repudiation is a challenge, especially if no trusted third party exists.
Blockchain promises to provide many of the required guarantees, which is why it has been used in many non-repudiable data exchange protocols.
Specifically, some authors propose to append data to blockchain to achieve non-repudiation of receipt.

In this position paper, we show that this approach is insufficient.
While appending data to blockchain can guarantee non-repudiation of origin in some cases, it is not sufficient for non-repudiation of receipt.
For confidential data, we find a catch-22 that makes it impossible.
For non-confidential data, meanwhile, plausible deniability remains.
We discuss potential solutions and suggest smart contracts as a promising approach.

\end{abstract}

\section{Introduction}

\begin{figure}[htbp]
	\centering
	\includegraphics{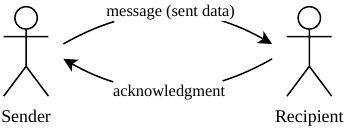}
	\caption{Terminology used in this paper. A \emph{sender} sends a \emph{message} to the \emph{recipient}, who is expected to respond with an \emph{acknowledgment}. Considering non-repudiation evidence in this simple example, the signed message could provide \emph{non-repudiation of origin} and the signed acknowledgment \emph{non-repudiation of receipt}.}
	\label{fig:terms}
\end{figure}

The Internet and peer-to-peer networks enable fully digital interactions.
For example, documents can be shared, messages exchanged, or even contracts signed, all digitally.
While the ease of use undoubtedly increases when digitalizing these use cases, the lack of physical interaction makes it easier for the involved parties to repudiate their involvement.
This can be problematic, especially when legal accountability is sought after.
For example, imagine the use case of certified email~\cite[e.g.,][]{ateniese2001tricert}, which could be useful when delivering legal papers.
The sender~(see also \autoref{fig:terms}) wants to ensure that they can prove the message was received, as legal obligations arise.
This is referred to as \emph{non-repudiation of receipt}, or NRR.
The recipient, meanwhile, has a diverging view: they want to be sure that they only acknowledge receipt if they received the expected message from a specific sender.
Should it be, e.g., lost in transit, they do not want to be held accountable.
This property is referred to as the \emph{fairness} of the protocol.
A fair protocol ensures that non-repudiation evidence is created if and only if the exchange was successful, and otherwise no valuable information is transferred~\cite[1608]{kremer2002intensive}.
To realize such digital delivery with non-repudiation, a trusted intermediary, also known as trusted third party (TTP), is useful.
The parties can send the message and the acknowledgment to the TTP, which only forwards them after having received both.
Yet, the required participation of a TTP has disadvantages.
For example, a TTP can become a bottleneck if large numbers of messages are sent.
Furthermore, it represents a single point of attack that could be, e.g., bribed or otherwise manipulated.
Therefore, approaches to non-repudiation without a TTP become relevant.

Most non-repudiation protocols we know of involve some form of TTP.
In fact, many authors consider non-repudiation to be impossible without a TTP~\cite[e.g.,][]{wang2006generic, kupccu2012usable, dziembowski2018fairswap};~\cite[see also][]{pagnia1999impossibility, garbinato2010impossibility}.
Recently, though, a promising area of research towards decentralized non-repudiable data exchange has emerged: delivering data via blockchain~\cite[see, e.g.,][]{mutpuigserver2018blockchain, marsalek2018sedicom, hinarejos2019solution, anand2019mirage, algoni2020p2p, zhang2021revocable, wang2021staged, chen2022blockchain, egala2022coviblock}.
Blockchain's inherent properties, notably immutability and availability, promise to ensure many of the required guarantees for non-repudiation.
Intuitively, the sender will not be able to repudiate their involvement, as the blockchain append of the payload is immutably recorded.
For the recipient, the core assumption is often that, due to blockchain's inherently public nature, the receipt of data is simply ``undeniable''~\cite[61]{zhang2021revocable} as they ``cannot deny having received'' it~\cite[31338]{hinarejos2019solution};~\cite[see also][16]{egala2022coviblock}.
Meaning, non-repudiation of receipt is assumed to be given.

In this paper, we show that this assumption does not hold.
First, when considering confidential data that need to be encrypted before sending, we find that the approach cannot provide any additional guarantees (\autoref{sec:issues:data-protection}).
Even for publicly shareable data, though, we find that plausible deniability for the recipient remains, which precludes non-repudiation (\autoref{sec:issues:plausible-deniability}).
To provide some constructive value as well, we then give an overview of two alternative approaches and discuss their merits (\autoref{sec:solutions}).

\section{First Issue: Confidentiality Catch-22}
\label{sec:issues:data-protection}

In this section, we show:

\begin{theorem} \label{thm:confidentiality-catch-22}
	Delivering confidential data via blockchain cannot guarantee non-repudiation of receipt.
\end{theorem}

To start with, blockchain can only give guarantees of data security and availability if a large number of unrelated parties participate in the network.
Consequently, these parties can also read any data appended to the blockchain.
We simplify this as:

\begin{lemma}
	Appending data to blockchain makes them publicly readable.
\end{lemma}

The public nature of blockchain is useful for non-repudiation, but can become an issue depending on the sent data.
On the one hand, data protection laws such as the European GDPR include confidentiality requirements that forbid public accessibility of personal data~\cite{eu2016gdpr} and require it to be deletable after rightful requests~\cite[see][]{pagallo2018chronicle}.
A typical motivating example are health data~\cite[e.g.,][]{algoni2020p2p, egala2022coviblock}, which are considered sensitive.
Thus, appending such data without technical protections, i.e. encryption, to blockchain may be illegal.
These protections must make the data unreadable for third parties, to make reidentification of data subjects impossible.
On the other hand, even non-personal data can still represent value, as in the case of digital media~\cite[e.g.,][]{onieva2007integration} or valuable business data~\cite[e.g.,][]{genes2022data}.
The sender therefore may not want arbitrary blockchain readers to be able to access these data.
Again, technical protections become necessary.
Therefore, we note:

\begin{lemma} \label{lma:confidential-data-unreadable}
	Confidential data may only be appended to blockchain if they are transformed to be unreadable for third parties.
\end{lemma}

\begin{lemma}
	To make data unreadable while preserving their value for the recipient, they are encrypted with a key that can be delivered separately.
\end{lemma}

Yet, if the data can only be read by decrypting them with a key, the same problem of repudiability returns, just for said key.
Meaning, the recipient can claim to not have received the \emph{key} (instead of the datum), rendering the approach ineffectual.
As the data stored on the blockchain are unreadable without the key~(see \autoref{lma:confidential-data-unreadable}), the participation in the blockchain network alone does not suffice to ``receive'' them.
Therefore, a catch-22:
Either confidential data are stored publicly on blockchain, which must not happen, or non-repudiation of receipt cannot be guaranteed without additional measures.
\hfill
\qed

\section{Second Issue: Plausible Deniability}
\label{sec:issues:plausible-deniability}

In our view, most use cases for non-repudiation implicate some form of confidentiality requirements.
Still, let us assume that there are applications using non-confidential data that nonetheless require non-repudiation.
With non-confidential data, we refer to all data that can become public without issue and can therefore be shared with arbitrary parties.

Consequently, in this section, we show:

\begin{theorem} \label{thm:plausible-deniability}
	Delivering non-confidential data via blockchain cannot guarantee non-repudiation of receipt.
\end{theorem}

In case of non-confidential data, the confidentiality catch-22 does not apply, as the data do not need to be protected.
Accordingly, they may be appended as plaintext to the blockchain.
Given this, previous works assume non-repudiation to be given due to blockchain's availability (see above).
The core of the argument is that the availability of blockchain means that (re)downloading the data is always possible.
Therefore, the transfer can be considered to be always successful~\cite[see, e.g.,][75]{zhang2021revocable}.
Yet, non-repudiation of receipt can only be assumed if no plausible deniability for the recipient remains.
We can relatively simply construct a realistic scenario for plausible deniability, though---the recipient can always claim they disconnected from the network, even after successfully receiving the data.
For example, their computer may have broken down, their internet connection blocked, or their power failed.
The sender will be unable to refute such claims.
Therefore, non-repudiation of receipt cannot be guaranteed.
\hfill
\qed

Let us go a step further, though.
It seems that in many of the related works, an implicit assumption is not only that it is always \emph{possible} for the recipient to download the data, but that is in fact \emph{expected} of them.
Blockchain does seem to enable this.
If the receipt is unsuccessful, e.g. due to a network error as described above, the recipient is always able to retry downloading the data by syncing with the network.
We can therefore deduce:

\begin{lemma} \label{lma:nrr-requires-retries}
	Non-repudiation of receipt for data delivered via blockchain depends on the requirement toward the recipient to retry failed transfers until they succeed.
\end{lemma}

Yet, in our view, this requirement does not hold in practice.
In fact, the recipient \emph{should} always be able to deny receipt.
Consider that not all messages are sent consensually.
A malicious actor may send a message with, e.g., illegal material.
If the recipient would be \emph{required} to download the message (see \autoref{lma:nrr-requires-retries}), they may be liable to prosecution due to possession of illegal material.
Therefore, this requirement cannot be enforced, making it unfit to guarantee non-repudiation of receipt.
\hfill
\qed

\section{Solutions}
\label{sec:solutions}

In the following, we give an overview of alternative solution approaches to non-repudiation utilizing blockchain.
We argue that more complex delivery protocols are not the solution and instead point to smart contracts as a promising technology for non-repudiation.

\subsection{Staged Data Delivery via Blockchain}

Promising to solve the conflict of delivering confidential data, staged protocols have been proposed~\cite[e.g.,][]{wang2021staged, chen2022blockchain}.
Here, the shared datum is split up into multiple parts.
We can generalize this solution as splitting the payload up into two halves, e.g. the encrypted data and the encryption key, as increasing the number of parts arbitrarily does not change the provided guarantees.
The core of the idea is that instead of sharing the complete payload via the blockchain, only one half is appended.
First, the data owner sends an unreadable half of the data directly.
Then, the consumer acknowledges the receipt.
Finally, the owner shares the second half of the data via the blockchain network with the consumer~\cite{wang2021staged}.
The order may also be reversed~\cite[e.g.,][]{chen2022blockchain}, which is functionally identical, though.
Thereby, the problem of having to store personally identifiable data publicly does not apply if only an unreadable part of the data is added to the blockchain.
Critically, though, the process suffers from exactly the same plausible deniability issue noted above (see \autoref{thm:plausible-deniability}), as the receipt of the second half of the data again depends on the ``undeniable'' delivery via the blockchain.
As the recipient cannot decipher the datum without receiving the last part, they can always repudiate the receipt of the full datum.
Therefore, non-repudiation of receipt cannot be guaranteed.

\subsection{Smart Contracts as the Intermediary}

We find that approaches based, fundamentally, on a blockchain append may not suffice for non-repudiation of receipt.
Alternative approaches have been proposed, utilizing a smart contract as the intermediary~\cite[e.g.,][]{dziembowski2018fairswap, eckey2020optiswap}.
While this can be considered a TTP~\cite[see][Sec.~6.1]{zieglmeier2023decentralized}, we consider it a weaker notion of TTP, as the behavior of the smart contract is immutable, transparent, and auditable.
Utilizing a smart contract has an important advantage compared to other approaches, in that existing non-repudiation protocols requiring a TTP can be used.
For example, smart contracts can be used to implement arbitrated exchange~\cite[e.g.,][]{dziembowski2018fairswap} or optimistic fair exchange~\cite[e.g.,][]{eckey2020optiswap}.
Accordingly, the protocols can benefit from the extensive existing security analysis for fair exchange with a TTP.
Importantly, though, security vulnerabilities in smart contracts exist~\cite[see][]{chen2020survey} and represent a unique attack vector of this approach.

\section{Discussion and Implications}

Appending data to blockchain is commonly assumed to be sufficient for non-repudiation of receipt in a data exchange.
We show that this assumption does not hold in practice, meaning that additional steps are required to achieve non-repudiation.
This should not be understood as a call to develop more complex protocols that are based, fundamentally, on a blockchain append.
Instead, we argue that smart contracts may be the more promising way forward.

The assumption that blockchain can guarantee non-repudiation of receipt ``out of the box'' can be misleading or, depending on the use case, wrong.
Approaches that were developed based on this assumption can still be relevant, though.
While we show that they do not provide non-repudiation of receipt, their other useful properties remain.
For use cases such as contract signing, for example, non-repudiation of origin may suffice.

We hope to inspire future research that closes this gap, with a solution that truly makes the receipt of data ``undeniable.''

\subsubsection{Acknowledgements}
We thank Jenny Lin for screening the literature to compile examples and for many fruitful discussions that helped us to refine the argumentation.

\printbibliography{}
\end{document}